\documentclass[12pt,openany,oneside]{article}
\usepackage{axodraw}
\usepackage{psfig,color,epsfig}
\usepackage{epsf}
\usepackage{graphicx}
\usepackage{latexsym}
\newcommand{\ee}{\end{equation}}
\newcommand{\be}{\begin{equation}}
\newcommand{\ea}{\end{array}}
\newcommand{\bqa}{\begin{eqnarray}}
\newcommand{\eqa}{\end{eqnarray}}
\begin{document}
\begin{center}
{\Large\bf Canonical Formalism for Lagrangians of Maximal
Nonlocality}
\\[10mm]
{\sc Hanghui Chen and H.~Q.~Zheng}
\\[2mm]
{\it  Department of Physics, Peking University, Beijing 100871,
P.~R.~China}
\\[5mm]
\today
\begin{abstract}

A canonical formalism for Lagrangians of maximal nonlocality is
established. The method is based on the familiar Legendre
transformation to a new function which can be derived from the
maximally nonlocal Lagrangian. The corresponding canonical
equations are derived through the standard procedure in local
theory and appear much like those local ones, though the
implication of the equations is largely expanded.
\end{abstract}
\end{center}

\section{Introduction}

It has been long acknowledged that a physically acceptable Hamiltonian is
bounded below, otherwise many fundamental concepts in physics would be
challenged, including the statistical interpretation of wave function and
the existence of ground state. A Hamiltonian which is unbounded below would
lead to negative probability and unending transition from high excited state
to lower and lower energies without bound, thus making the system unstable.
The criterion that a Hamiltonian should be bounded below determines which
candidate Lagrangians can survive, on the most fundamental level.

An important theorem relevant to the criterion above is that a
Lagrangian depends only upon the zeroth and first time
derivatives, which is obtained by the 19$^{th}$ century physicist
Ostrogradski~\cite{1}. Allowing higher time derivatives always
leads to a Hamiltonian which is not bounded below and the
instability of the system. When nonlocality is introduced into
Lagrangians, many new motion solutions and features emerge, some
of which appear miraculous from the context of a local theory. A
meaningful consequence is obtained in Ref.~\cite{2} that a
Lagrangian with nonlocality of finite extent, which can be
represented as the limits of higher derivatives, inherit the full
Ostrogradskian instability. However, three Lagrangians of maximal
nonlocality are raised in Ref.~\cite{3} to indicate that the
Hamiltonian of a maximally nonlocal Lagrangian can be bounded
below, so there is no generic instability of the Ostrogradskian
type. In order to gain a deeper understanding of maximally
nonlocal Lagrangians and apply them to quantum theory, a canonical
formalism seems a necessary intermediate step towards
quantization.

The purpose of this paper is to construct a canonical formalism
for maximally nonlocal Lagrangians and the corresponding canonical
equations. The paper is organized as follows: Section~\ref{sec2}
is devoted to the construction of the new canonical formalism and
canonical equations. This formalism is applied in Sec.~\ref{sec3}
to the three examples raised in Ref.~\cite{3} to indicate that a
maximally nonlocal Lagrangian can survive due to a bounded below
Hamiltonian and to specify some important and fundamental concepts
which are not emphasized in Ref.~\cite{3}. Another example is also
raised in Sec.~\ref{sec3} to furnish a deeper understanding of the
criterion concerning the stability of the system. The conclusions
comprise Sec.~\ref{sec4}.

\section{Construction of Canonical Formalism and
Canonical Equations}\label{sec2}

A maximally nonlocal Lagrangian is defined as one which
potentially depends upon the dynamical variables from time $ - T$
to time $T$, where $T$ is a large positive number approaching $ +
\infty $. The new Lagrangian is written in an explicit form as:
\begin{equation}
\label{eq1} L = L\left( {q,\dot {q},t,F(q,\dot {q})} \right)\ ,
\end{equation}
where $F(q,\dot {q})$ has a general form as:
\begin{equation}
\label{eq2} F(q,\dot {q}) = \mathop {\lim }\limits_{T \to \infty }
\int\limits_{ - T}^T {f(q,\dot {q},t)dt}\ .
\end{equation}
When dealing with such a Lagrangian, we have to make some
modifications to Hamilton's principle. In standard local cases,
the arbitrary integration bounds usually cause no trouble as far
as the Euler-Lagrangian equations are concerned. However, the
integration bounds play such an important role in nonlocal action
that different motion equations can be derived from different
forms of the integration bounds~\cite{4}. Hence, we assume the
integration bounds in Hamilton's principle to be in similar forms
as those in Eq.~(\ref{eq2}), i.e.
\begin{equation}
\label{eq3} S = \mathop {\lim }\limits_{T \to \infty }
\int\limits_{ - T}^T {L\left( {q,\dot {q},t,F(q,\dot {q})}
\right)dt}\ .
\end{equation}
Moreover, we make an agreement that the integration bound $T$ in
Eq.~(\ref{eq2}) and Eq.~(\ref{eq3}) be considered as a constant
and only when the equations of motion are obtained will $T$
approach $ + \infty $. The purpose of such a procedure is to avoid
some possible divergent problems~\cite{3}.

According to the standard procedure, the variation of the action
integral $S$ in Eq.~(\ref{eq3}) is:
\[
\delta S = \mathop {\lim }\limits_{T \to \infty } \int\limits_{ -
T}^T {\left[ {\frac{\partial L}{\partial q}\delta q +
\frac{\partial L}{\partial \dot {q}}\delta \dot {q} +
\frac{\partial L}{\partial F}\int\limits_{ - T}^T {\left(
{\frac{\partial f}{\partial q}\delta q + \frac{\partial
f}{\partial \dot {q}}\delta \dot {q}} \right)dt} } \right]} dt
\]
\begin{equation}
\label{eq4}
 = \mathop {\lim }\limits_{T \to \infty } \int\limits_{ - T}^T {\left[
{\frac{\partial L}{\partial q}\delta q + \frac{\partial
L}{\partial \dot {q}}\delta \dot {q} + \left( {\int\limits_{ -
T}^T {\frac{\partial L}{\partial F}dt} } \right)\left(
{\frac{\partial f}{\partial q}\delta q + \frac{\partial
f}{\partial \dot {q}}\delta \dot {q}} \right)} \right]} dt\ .
\end{equation}
Considering the boundary conditions
$$\mathop {\lim }\limits_{T \to
\infty } \delta q(T) = \mathop {\lim }\limits_{T \to \infty }
\delta q( - T) = 0\ ,$$
 we can obtain
$$\delta S = \mathop {\lim
}\limits_{T \to \infty } \int\limits_{ - T}^T {\left[
{\frac{\partial L}{\partial q} - \frac{d}{dt}\left(
{\frac{\partial L}{\partial \dot {q}}} \right) + \left(
{\int\limits_{ - T}^T {\frac{\partial L}{\partial F}dt} }
\right)\left( {\frac{\partial f}{\partial q} - \frac{d}{dt}\left(
{\frac{\partial f}{\partial \dot {q}}} \right)} \right)} \right]}
\delta qdt\ .$$
 The new equation of motion is derived from:
\begin{equation}
\label{eq5} \delta S = 0,\forall \delta q\ .
\end{equation}
Hence, it leads to:
\begin{equation}
\label{eq6} \frac{\partial L}{\partial q} - \frac{d}{dt}\left(
{\frac{\partial L}{\partial \dot {q}}} \right) + \left( {\mathop
{\lim }\limits_{T \to \infty } \int\limits_{ - T}^T
{\frac{\partial L}{\partial F}dt} } \right)\left( {\frac{\partial
f}{\partial q} - \frac{d}{dt}\left( {\frac{\partial f}{\partial
\dot {q}}} \right)} \right) = 0\ .
\end{equation}
We define a new function $L_c $ as:
\begin{equation}
\label{eq7} L_c = L + \left( {\mathop {\lim }\limits_{T \to \infty
} \int\limits_{ - T}^T {\frac{\partial L}{\partial F}dt} }
\right)\cdot f\ .
\end{equation}
Note that $\left( {\mathop {\lim }\limits_{T \to \infty }
\int\limits_{ - T}^T {\frac{\partial L}{\partial F}dt} } \right)$
behaves like a constant as far as $\frac{\partial }{\partial q}$
and $\frac{\partial }{\partial \dot {q}}$ are concerned. The new
equation of motion can be reduced to a simple form, which is
similar to the local case:
\begin{equation}
\label{eq8} \frac{\partial L_c }{\partial q} - \frac{d}{dt}\left(
{\frac{\partial L_c }{\partial \dot {q}}} \right) = 0\ .
\end{equation}
Before the canonical formalism is constructed, two important
conserved quantities are discussed regarding the motion of maximal
nonlocality. The first conserved quantity results from the
homogeneity of time~\cite{5}. By virtue of this homogeneity, the
properties of a system are unchanged by any displacement in time.
Consider an infinitesimal displacement $\varepsilon $ in time and
the corresponding change in $L$, the generalized coordinates and
velocities remaining fixed, is:
\[
\delta L = \frac{\partial L}{\partial t}\delta t + \frac{\partial
L}{\partial F}\mathop {\lim }\limits_{T \to \infty } \int\limits_{ - T}^T
{\frac{\partial f}{\partial t}} \delta tdt
\]
\begin{equation}
\label{eq9}
\quad
 = \varepsilon \cdot \left( {\frac{\partial L}{\partial t} + \frac{\partial
L}{\partial F}\mathop {\lim }\limits_{T \to \infty } \int\limits_{
- T}^T {\frac{\partial f}{\partial t}} dt} \right)\ .
\end{equation}
Since $\varepsilon $ is arbitrary, the condition $\delta L = 0$ is
equivalent to:
\begin{equation}
\label{eq10} \frac{\partial L}{\partial t} + \frac{\partial
L}{\partial F}\mathop {\lim }\limits_{T \to \infty } \int\limits_{
- T}^T {\frac{\partial f}{\partial t}} dt = 0\ .
\end{equation}
However, as far as the motion of the maximal nonlocality is
concerned, we discuss a stronger condition to satisfy the
homogeneity of time, i.e. \be \frac{\partial L}{\partial t} = 0
\,\,\,\mathrm{and}\,\,\, \frac{\partial f}{\partial t} = 0\ .
\label{12} \ee
 Under the
condition Eq.~(\ref{12}) and noting the definition of $L_c $ in
Eq.~(\ref{eq7}), we can obtain:
\begin{equation}
\label{eq11} \frac{\partial L_c }{\partial t} = 0\ .
\end{equation}
The total time derivative of the new function $L_c $ can therefore
be written as,
\begin{equation}
\label{eq12} \frac{dL_c }{dt} = \frac{\partial L_c }{\partial
q}\dot {q} + \frac{\partial L_c }{\partial \dot {q}}\ddot {q}\ .
\end{equation}
Replacing $\frac{\partial L_c }{\partial q}$, in accordance with
the new equation of motion, by $\frac{d}{dt}\left( {\frac{\partial
L_c }{\partial \dot {q}}} \right)$, we obtain:
 \be\label{eq13}
\frac{dL_c }{dt} = \dot {q}\frac{d}{dt}\left( {\frac{\partial L_c
}{\partial \dot {q}}} \right) + \frac{\partial L_c }{\partial \dot
{q}}\ddot {q}
%
%
 = \frac{d}{dt}\left( {\dot {q}\frac{\partial L_c }{\partial \dot {q}}}
\right)\ ,
\end{equation}
or
\begin{equation}
\label{eq14} \frac{d}{dt}\left( {\dot {q}\frac{\partial L_c
}{\partial \dot {q}} - L_c } \right) = 0\ .
\end{equation}
We therefore define the energy of the system as:
\begin{equation}
\label{eq15} E = \dot {q}\frac{\partial L_c }{\partial \dot {q}} -
L_c\ ,
\end{equation}
which, under the condition Eq.~(\ref{12})\footnote{A more general
condition to obtain the first conserved quantity from the
homogeneity of time is based on Eq.~(\ref{eq10}):
\begin{eqnarray*}
\frac{\partial L_c}{\partial t}&=&\frac{\partial L}{\partial
t}+\left(\lim_{T\to \infty}\int^T_{-T}\frac{\partial L}{\partial
F}dt\right)\cdot\frac{\partial f}{\partial t}\nonumber\\
&=&\left(\lim_{T\to \infty}\int^T_{-T}\frac{\partial L}{\partial
F}dt\right)\cdot\frac{\partial f}{\partial t}-\frac{\partial
L}{\partial F}\cdot\left(\lim_{T\to
\infty}\int^T_{-T}\frac{\partial f}{\partial t}dt\right)\ .
\end{eqnarray*}
 }, is a conserved quantity during the motion of
the maximal nonlocality.

The second conserved quantity follows from the homogeneity of
space~\cite{5}. By virtue of this homogeneity, the properties of a
system are unchanged by any parallel displacement in space.
Consider an infinitesimal displacement $\varepsilon $ in space and
the corresponding change in $L$, the generalized velocities
remaining fixed, is:
\[
\delta L = \frac{\partial L}{\partial q}\delta q + \frac{\partial
L}{\partial F}\mathop {\lim }\limits_{T \to \infty } \int\limits_{ - T}^T
{\frac{\partial f}{\partial q}} \delta qdt
\]
\begin{equation}
\label{eq16}
\quad
 = \varepsilon \cdot \left( {\frac{\partial L}{\partial q} + \frac{\partial
L}{\partial F}\mathop {\lim }\limits_{T \to \infty } \int\limits_{
- T}^T {\frac{\partial f}{\partial q}} dt} \right)\ .
\end{equation}
Since $\varepsilon $ is arbitrary, the condition $\delta L = 0$ is
equivalent to:
\begin{equation}
\label{eq17} \frac{\partial L}{\partial q} + \frac{\partial
L}{\partial F}\mathop {\lim }\limits_{T \to \infty } \int\limits_{
- T}^T {\frac{\partial f}{\partial q}} dt = 0\ .
\end{equation}
Similarly, as far as the motion of the maximal nonlocality is
concerned, we discuss a stronger condition to satisfy the
homogeneity of space, i.e.
 \be \frac{\partial L}{\partial q} =
0\,\,\, \mathrm{and}\,\,\, \frac{\partial f}{\partial q} = 0\
.\label{20}\ee
 Under the
condition Eq.~(\ref{20}) and noting the definition of $L_c $ in
Eq.~(\ref{eq7}), we can obtain:
\begin{equation}
\label{eq18} \frac{\partial L_c }{\partial q} = 0\ .
\end{equation}
By the new equation of motion Eq.~(\ref{eq8}), Eq.~(\ref{eq18}) is
equivalent to:
\begin{equation}
\label{eq19} \frac{d}{dt}\left( {\frac{\partial L_c }{\partial
\dot {q}}} \right) = 0\ .
\end{equation}
We define the canonical momentum of the system as:
\begin{equation}
\label{eq20} p^\ast = \frac{\partial L_c }{\partial \dot {q}}\ ,
\end{equation}
which, under the condition Eq.~(\ref{20})\footnote{A more general
condition to obtain the second conserved quantity from the
homogeneity of space is based on Eq.~(\ref{eq17}):
\begin{eqnarray*}
\frac{\partial L_c}{\partial q}&=&\frac{\partial L}{\partial
q}+\left(\mathop{\lim}_{T\to \infty}\int^T_{-T}\frac{\partial
L}{\partial
F}dt\right)\cdot\frac{\partial f}{\partial q}\nonumber\\
&=&\left(\lim_{T\to \infty}\int^T_{-T}\frac{\partial L}{\partial
F}dt\right)\cdot\frac{\partial f}{\partial q}-\frac{\partial
L}{\partial F}\cdot\left(\lim_{T\to
\infty}\int^T_{-T}\frac{\partial f}{\partial q}dt\right)\ .
\end{eqnarray*}
 }, is a
conserved quantity during the motion of the maximal nonlocality.

By the Legendre transformation, we change the variables from $(q,\dot
{q},t)$ to $(q,p^\ast ,t)$ and define the new Hamiltonian as:
\begin{equation}
\label{eq21} H^\ast \left( {q,p^\ast ,t,G(q,p^\ast )} \right)
\equiv p^\ast q - L_c \left( {q,\dot {q},t,F(q,\dot {q})} \right)\
.
\end{equation}
Note that as the Hamiltonian depends on $q,p^\ast $and $t$ only,
the generalized velocity $\dot {q}$ has to be converted into the
function of $q,p^\ast ,t$ and $G(q,p^\ast )$, where $G(q,p^\ast )$
has a general form as:
\begin{equation}
\label{eq22} G(q,p^\ast ) = \mathop {\lim }\limits_{T \to \infty }
\int\limits_{ - T}^T {g(q,p^\ast ,t)dt}\ .
\end{equation}
A general method of converting $\dot {q}$ to $\dot {q} = \dot
{q}\left( {q,p^\ast ,t,G(q,p^\ast )} \right)$ is furnished below.

By the definition of $p^\ast $ in Eq.~(\ref{eq20}), we can obtain
a general expression of the new canonical momentum:
\begin{equation}
\label{eq23} p^\ast = p^\ast \left( {q,\dot {q},t,F(q,\dot {q})}
\right)\ ,
\end{equation}
where $F(q,\dot {q})$ has a general form as $\mathop {\lim
}\limits_{T \to \infty } \int\limits_{ - T}^T {f(q,\dot {q},t)dt}
$. Let $F(q,\dot {q})$ be $\gamma $ and assume that $\dot {q}$ can
be solved from Eq.~(\ref{eq23}), i.e.
\begin{equation}
\label{eq24} \dot {q} = \dot {q}\left( {q,p^\ast ,t,\gamma }
\right)\ ,
\end{equation}
then substitute Eq.~(\ref{eq24}) into the general form of
$F(q,\dot {q})$ and we obtain:
\begin{equation}
\label{eq25} \mathop {\lim }\limits_{T \to \infty } \int\limits_{
- T}^T {f\left( {q,\dot {q}(q,p^\ast ,t,\gamma ),t} \right)dt =
\gamma }\ .
\end{equation}
This is the algebra equation for $\gamma $. The solution of
Eq.~(\ref{eq25}) is denoted as $\gamma _0 $, which is the function
of $G(q,p^\ast )$, where $G(q,p^\ast )$ has a general form as
$\mathop {\lim }\limits_{T \to \infty } \int\limits_{ - T}^T
{g(q,p^\ast ,t)dt} $. Substitute the solution $\gamma _0 \left(
{G(q,p^\ast )} \right)$ back into Eq.~(\ref{eq24}) we obtain:
\begin{equation}
\label{eq26} \dot {q} = \dot {q}\left( {q,p^\ast ,t,\gamma _{_0 }
\left( {G(q,p^\ast ,t)} \right)} \right)\ .
\end{equation}
Considered as a function of $q,p^\ast $ and $t$ only, the
differential of $H^\ast $is given by:
\begin{equation}
\label{eq27} dH^\ast = \frac{\partial H^\ast }{\partial q}dq +
\frac{\partial H^\ast }{\partial p^\ast }dp^\ast + \frac{\partial
H^\ast }{\partial t}dt\ .
\end{equation}
It is emphasized again that as far as $\frac{\partial }{\partial
q}$ and $\frac{\partial }{\partial p^\ast }$ are concerned,
$G(q,p^\ast ) = \mathop {\lim }\limits_{T \to \infty }
\int\limits_{ - T}^T {g(q,p^\ast ,t)dt} $ is considered as a
constant. But from the defining equation (\ref{eq21}), we can also
write:
\begin{equation}
\label{eq28} dH^\ast = d\left( {p^\ast \dot {q} - L_c } \right) =
\dot {p}^\ast dq + \dot {q}dp^\ast - dL_c\ .
\end{equation}
Considered as a function of $q,\dot {q}$ and $t$ only, the
differential of $L_c $is given by:
\begin{equation}
\label{eq29} dL_c = \frac{\partial L_c }{\partial q}dq +
\frac{\partial L_c }{\partial \dot {q}}d\dot {q} + \frac{\partial
L_c }{\partial t}dt\ .
\end{equation}
Noting the definition of $p^\ast $ (\ref{eq20}) and the equation
of motion Eq.~(\ref{eq8}), we can reduce Eq.~(\ref{eq29}) to a
simple form:
\begin{equation}
\label{eq30} dL_c = \dot {p}^\ast dq + p^\ast d\dot {q} +
\frac{\partial L_c }{\partial t}dt\ .
\end{equation}
Substituting Eq.~(\ref{eq30}) into Eq.~(\ref{eq28}), we can
obtain:
\begin{equation}
\label{eq31} dH^\ast = - \dot {p}^\ast dq + \dot {q}dp^\ast -
\frac{\partial L_c }{\partial t}dt\ .
\end{equation}
Comparison of Eq.~(\ref{eq31}) with Eq.~(\ref{eq27}) leads to the
new canonical equations of Hamiltonian:
\begin{equation}
\label{eq32}
\left\{ {\begin{array}{l}
 \frac{\partial H^\ast }{\partial p^\ast } = \dot {q}\ , \\
 \frac{\partial H^\ast }{\partial q} = - \dot {p}^\ast\ , \\
 \frac{\partial H^\ast }{\partial t} = - \frac{\partial L_c }{\partial t}\ , \\
 \end{array}} \right.
\end{equation}
which are similar to those of local cases.

\section{ A FEW EXAMPLES}\label{sec3}

Applications of the new canonical formalism and canonical
equations are illustrated in this section, where $\mathop {\lim
}\limits_{T \to \infty } $ will be omitted for the sake of
conciseness and convenience. The implication of the integration
bound accords with the agreement in Sec.~\ref{sec2}.
 The new canonical formalism and canonical equations are first
applied to the three Lagrangians in Ref.~\cite{3}.

Consider the first Lagrangian: \be L = \frac{1}{2}m\dot {q}^2 -
\frac{1}{4}m\omega _0^2 q^2\int\limits_{ - T}^T {\frac{q^2}{l_0^2
}} \frac{dt}{T}\ ,\label{36} \ee
 where $F(q,\dot {q}) = \int\limits_{ - T}^T {\frac{q^2}{l_0^2 }}
\frac{dt}{T}$ and $f(q,\dot {q},t) = \frac{q^2}{Tl_0^2 }\ .$ The
new function $L_c $ can be derived from Eq.~(\ref{eq7}):
\begin{equation}
\label{eq33} L_c = L + \left( {\int\limits_{ - T}^T
{\frac{\partial L}{\partial F}dt} } \right) \cdot f =
\frac{1}{2}m\dot {q}^2 - \frac{1}{2}m\omega _0^2 q^2\int\limits_{
- T}^T {\frac{q^2}{l_0^2 }\frac{dt}{T}}\ .
\end{equation}
The energy of the system is:
\begin{equation}
\label{eq34} E = \dot {q}\frac{\partial L_c }{\partial \dot {q}} -
L_c = \frac{1}{2}m\dot {q}^2 + \frac{1}{2}m\omega _0^2
q^2\int\limits_{ - T}^T {\frac{q^2}{l_0^2 }\frac{dt}{T}}\ ,
\end{equation}
which is conserved as the condition Eq.~(12) is satisfied. The
canonical momentum is:
\begin{equation}
\label{eq35} p^\ast = \frac{\partial L_c }{\partial \dot {q}} =
m\dot {q}\ .
\end{equation}
Convert $\dot {q}$ to:
\begin{equation}
\label{eq36} \dot {q} = \frac{p^\ast }{m}\ ,
\end{equation}
we can obtain the new Hamiltonian from Eq.~(\ref{eq33}) and
Eq.~(\ref{eq36}):
\begin{equation}
\label{eq37} H^\ast = p^\ast \dot {q} - L_c = \frac{(p^\ast
)^2}{2m} + \frac{1}{2}m\omega _0^2 q^2\int\limits_{ - T}^T
{\frac{q^2}{l_0^2 }\frac{dt}{T}}\ .
\end{equation}
By the new canonical equation (\ref{eq32}), the equation of motion
reads:
\begin{equation}
\label{eq38} \ddot {q} + \omega ^2q = 0\ ,
\end{equation}
 where
  \be\omega ^2 = \omega _0^2 \int\limits_{ - T}^T
{\frac{q^2}{l_0^2 }} \frac{dt}{T}\ .\label{43}\ee
 It is in
accordance with Eq.~(10-11) in Ref.~[3].

Consider the second Lagrangian:
\begin{equation}
\label{eq39} L = \frac{1}{2}m\dot {q}^2 - \frac{1}{4}m\omega _0^2
q^2 - \frac{1}{16}m\omega _0^2 l_0^2 \sin \left(
{\frac{2}{T}\int\limits_{ - T}^T {\frac{q^2}{l_0^2 }} dt} \right)\
.
\end{equation}
where $F(q,\dot {q}) = 2\int\limits_{ - T}^T {\frac{q^2}{l_0^2 }}
\frac{dt}{T}$ and $f(q,\dot {q},t) = \frac{2q^2}{Tl_0^2 }$.
Repeating the same procedure, we have:
\begin{equation}
\label{eq40} L_c = \frac{1}{2}m\dot {q}^2 - \frac{1}{2}m\omega
_0^2 q^2\cos ^2\left( {2\int\limits_{ - T}^T {\frac{q^2}{l_0^2 }}
\frac{dt}{T}} \right) - \frac{1}{16}m\omega _0^2 l_0^2 \sin \left(
{2\int\limits_{ - T}^T {\frac{q^2}{l_0^2 }} \frac{dt}{T}} \right)\
.
\end{equation}
The energy of the system, which is conserved as the condition
Eq.~(\ref{12}) is satisfied, is:
\begin{equation}
\label{eq41} E = \frac{1}{2}m\dot {q}^2 + \frac{1}{2}m\omega _0^2
q^2\cos ^2\left( {2\int\limits_{ - T}^T {\frac{q^2}{l_0^2 }}
\frac{dt}{T}} \right) + \frac{1}{16}m\omega _0^2 l_0^2 \sin \left(
{2\int\limits_{ - T}^T {\frac{q^2}{l_0^2 }} \frac{dt}{T}} \right)\
,
\end{equation}
which accords with the energy defined in Ref.~[3] except for the
last added term on the $r.$$h.$$s.$ of the above equation.
  The
canonical momentum is:
\begin{equation}
\label{eq42} p^\ast = \frac{\partial L_c }{\partial \dot {q}} =
m\dot {q}\ ,
\end{equation}
\noindent or
\begin{equation}
\label{eq43} \dot {q} = \frac{p^\ast }{m}\ .
\end{equation}
The Hamiltonian is:
\begin{equation}
\label{eq44} H^\ast = \frac{(p^\ast )^2}{2m} + \frac{1}{2}m\omega
_0^2 q^2\cos ^2\left( {2\int\limits_{ - T}^T {\frac{q^2}{l_0^2 }}
\frac{dt}{T}} \right) + \frac{1}{16}m\omega _0^2 l_0^2 \sin \left(
{2\int\limits_{ - T}^T {\frac{q^2}{l_0^2 }} \frac{dt}{T}} \right)\
,
\end{equation}
and the equation of motion follows:
\begin{equation}
\label{eq45} \ddot {q} + \omega ^2q = 0\ ,
\end{equation}
\noindent where
 \be
\omega ^2 = \omega _0^2 \cos ^2\left( {\int\limits_{ - T}^T
{\frac{q^2}{l_0^2 }} \frac{dt}{T}} \right) \ .\label{51} \ee
 It is in
accordance with Eq.~(20-21) in Ref.~\cite{3}.

Now consider the third Lagrangian whose consequences will be a
little different from those in Ref.~[3].
\begin{equation}
\label{eq46} L = - \frac{1}{4}m\omega _0^2 l_0^2 \exp \left( { -
\int\limits_{ - T}^T {\frac{\dot {q}^2}{\omega _0^2 l_0^2
}\frac{dt}{T}} } \right) - \frac{1}{2}m\omega _0^2 q^2\ ,
\end{equation}
\noindent where $F(q,\dot {q}) = - \int\limits_{ - T}^T
{\frac{\dot {q}^2}{\omega _0^2 l_0^2 }\frac{dt}{T}} $ and
$f(q,\dot {q},t) = \frac{\dot {q}^2}{\omega _0^2 l_0^2 T}$. The
new function $L_c $ can be derived from Eq.~(\ref{eq7}):
\begin{equation}
\label{eq47} L_c = \frac{1}{2}m\dot {q}^2\exp \left( { -
\int\limits_{ - T}^T {\frac{\dot {q}^2}{\omega _0^2 l_0^2
}\frac{dt}{T}} } \right) - \frac{1}{2}m\omega _0^2 q^2 -
\frac{1}{4}m\omega _0^2 l_0^2 \exp \left( { - \int\limits_{ - T}^T
{\frac{\dot {q}^2}{\omega _0^2 l_0^2 }\frac{dt}{T}} } \right)\ .
\end{equation}
The energy of the system, which can be derived from
Eq.~(\ref{eq15}), follows:
\begin{equation}
\label{eq48} E = \frac{1}{2}m\dot {q}^2\exp \left( { -
\int\limits_{ - T}^T {\frac{\dot {q}^2}{\omega _0^2 l_0^2
}\frac{dt}{T}} } \right) + \frac{1}{2}m\omega _0^2 q^2 +
\frac{1}{4}m\omega _0^2 l_0^2 \exp \left( { - \int\limits_{ - T}^T
{\frac{\dot {q}^2}{\omega _0^2 l_0^2 }\frac{dt}{T}} } \right)\ ,
\end{equation}
\noindent which is conserved as the condition Eq.~(\ref{12}) is
satisfied. The canonical momentum is:
\begin{equation}
\label{eq49} p^\ast = m\dot {q}\exp \left( { - \int\limits_{ -
T}^T {\frac{\dot {q}^2}{\omega _0^2 l_0^2 }\frac{dt}{T}} }
\right)\ .
\end{equation}
According to the general method in Sec.~\ref{sec2}, let $ -
\int\limits_{ - T}^T {\frac{\dot {q}^2}{\omega _0^2 l_0^2
}\frac{dt}{T}} $ be $\gamma $:
\begin{equation}
\label{eq50}
p^\ast = m\dot {q}e^\gamma
\end{equation}
\noindent or
\begin{equation}
\label{eq51} \dot {q} = \frac{p^\ast }{m}e^{ - \gamma }\ .
\end{equation}
Substitute Eq.~(\ref{eq51}) into the explicit expression of
$\gamma $ we obtain:
\begin{equation}
\label{eq52}
 - \int\limits_{ - T}^T {\frac{1}{\omega _0^2 l_0^2 }\frac{(p^\ast
)^2}{m^2}e^{ - 2\gamma }\frac{dt}{T}} = \gamma\ .
\end{equation}
Now let $ - \int\limits_{ - T}^T {\frac{1}{\omega _0^2 l_0^2
}\frac{(p^\ast )^2}{m^2}\frac{dt}{T}} $ be $\alpha $ and
Eq.~(\ref{eq52}) is recast as:
\begin{equation}
\label{eq53} \gamma e^{2\gamma } = \alpha\ .
\end{equation}
The Eq.~(\ref{eq53}) is the algebra equation with respect to
$\gamma $, whose solution is determined by $\alpha $. Assume that
the solution to Eq.~(\ref{eq53}) is:
 \be\gamma _0 = \gamma _0
(\alpha )\ ,\label{60}
 \ee
 and substitute the solution Eq.~(\ref{60}) back into Eq.~(\ref{eq51}) we
 get:
 \begin{equation}
\label{eq54} \dot {q} = \frac{p^\ast }{m}e^{ - \gamma _0 (\alpha
)}\ ,
\end{equation}
\noindent where $\alpha = - \int\limits_{ - T}^T {\frac{1}{\omega
_0^2 l_0^2 }\frac{(p^\ast )^2}{m^2}\frac{dt}{T}} $. The new
Hamiltonian follows, noting that $\dot {q}$ has to appear in the
form of Eq.~(\ref{eq54}) in the Hamiltonian:
\begin{equation}
\label{eq55} H^\ast = \frac{(p^\ast )^2}{2m}e^{ - \gamma _0
(\alpha )} + \frac{1}{2}m\omega _0^2 q^2 + \frac{1}{4}m\omega _0^2
l_0^2 e^{\gamma _0 (\alpha )}\ .
\end{equation}
The equation of motion follows:
 \be \ddot {q} + \omega ^2q = 0\ ,\label{63}
  \ee
 \noindent where
  \be \omega ^2 = \omega _0^2 \exp
\left( {\int\limits_{ - T}^T {\frac{\dot {q}^2}{\omega _0^2 l_0^2
}\frac{dt}{T}} } \right)\ . \label{64}
\ee
 According to the method
in Ref.~[3], the general solution to Eq.~(\ref{63}) is:
\begin{equation}
\label{eq56} q(t) = q_0 \cos \omega t + \frac{\dot {q}_0 }{\omega
}\sin \omega t\ .
\end{equation}
Substituting Eq.~(\ref{eq56}) into Eq.~(\ref{64}), we can obtain
an algebra equation with respect to $\omega $ (Note that $T$
approaches $ + \infty $.):
\begin{equation}
\label{eq57} \omega ^2 = \omega _0^2 \exp \left( {\frac{\omega
^2q_0^2 + \dot {q}_0^2 }{\omega _0^2 l_0^2 }} \right)\ .
\end{equation}
One thing we should pay attention to is that from the solution
Eq.~(\ref{eq56}), it is taken for granted in Ref.~[3] that the
energy of the system is:
\begin{equation}
\label{eq58} E = \frac{1}{2}m\dot {q}^2 + \frac{1}{2}m\omega
^2q^2\ ,
\end{equation}
\noindent where $\omega $ is determined by Eq.~(\ref{eq57}).
However, substituting the general solution Eq.~(\ref{eq56}) into
$\exp \left( { - \int\limits_{ - T}^T {\frac{\dot {q}^2}{\omega
_0^2 l_0^2 }\frac{dt}{T}} } \right)$ in Eq.~(\ref{eq48}) and we
can obtain:
\begin{equation}
\label{eq59} E = \frac{1}{2}m\dot {q}^2\exp \left( { -
\frac{\omega ^2q_0^2 + \dot {q}_0^2 }{\omega _0^2 l_0^2 }} \right)
+ \frac{1}{2}m\omega _0^2 q^2 + \frac{1}{4}m\omega _0^2 l_0^2 \exp
\left( { - \frac{\omega ^2q_0^2 + \dot {q}_0^2 }{\omega _0^2 l_0^2
}} \right)\ .
\end{equation}
Note Eq.~(\ref{eq57}) and reduce Eq.~(\ref{eq59}) to:
\[
E = \frac{1}{2}m\dot {q}^2\left( {\frac{\omega _0^2 }{\omega ^2}} \right) +
\frac{1}{2}m\omega _0^2 q^2 + \frac{1}{4}m\omega _0^2 l_0^2 \left(
{\frac{\omega _0^2 }{\omega ^2}} \right)
\]
\begin{equation}
\label{eq60}
 = \left( {\frac{\omega _0^2 }{\omega ^2}} \right)\left( {\frac{1}{2}m\dot
{q}^2 + \frac{1}{2}m\omega ^2q^2 + \frac{1}{4}m\omega _0^2 l_0^2 }
\right)\ ,
\end{equation}
\noindent where $\omega $ is also determined by Eq.~(\ref{eq57}).

The Eq.~(\ref{eq60}) indicates that for every solution $\omega
^2$, the energy of the system is not the familiar form
Eq.~(\ref{eq58}) but rather a new form Eq.~(\ref{eq60}) which, if
the added constant $\frac{\omega _0^2 }{\omega
^2}\frac{1}{4}m\omega _0^2 l_0^2 $ is not taken into
consideration, is the product of Eq.~(\ref{eq58}) with
$\frac{\omega _0^2 }{\omega ^2}$. But $\omega ^2$ is not equal to
$\omega _0^2 $ except for specific initial value data to satisfy
Eq.~(\ref{eq57}), illustrating that Eq.~(\ref{eq60}) is not equal
to Eq.~(\ref{eq58}) on most occasions. The reason why the kinetic
energy is not equal to $\frac{1}{2}m\dot {q}^2$ is that the rule
determining the equations of motion is no longer Newton's second
law, which leads to a new definition of the kinetic energy.

The kinetic energy $T$ is defined in general to be:
\begin{equation}
\label{eq61} \mbox{ } dT \equiv \frac{dp^\ast }{dt}dq = \dot
{q}dp^\ast\ ,
\end{equation}
\noindent where $p^\ast $ is the new canonical momentum defined by
Eq.~(\ref{eq20}). Applying Eq.~(\ref{eq61}) to the previous
example, we can obtain the new kinetic energy from
Eq.~(\ref{eq49}):
\[
dT = \dot {q}dp^\ast = \dot {q}d\left( {m\dot {q}\exp \left( { -
\int\limits_{ - T}^T {\frac{\dot {q}^2}{\omega _0^2 l_0^2 }\frac{dt}{T}} }
\right)} \right)
\]
\begin{equation}
\label{eq62}
 = d\left( {\frac{1}{2}m\dot {q}^2\exp \left( { - \int\limits_{ - T}^T
{\frac{\dot {q}^2}{\omega _0^2 l_0^2 }\frac{dt}{T}} } \right)}
\right)\ .
\end{equation}
Substituting the general solution Eq.~(\ref{eq56}) into
Eq.~(\ref{eq62}) and noting the algebra equation Eq.~(\ref{eq57}),
we can obtain that for every solution $\omega ^2$, the new kinetic
energy is:
\begin{equation}
\label{eq63} dT = d\left( {\frac{1}{2}m\dot {q}^2\frac{\omega _0^2
}{\omega ^2}} \right)\ ,
\end{equation}
\noindent which accords with Eq.~(\ref{eq60}). However, the
difference in Eq.~(\ref{eq58}) and Eq.~(\ref{eq60}) does not
disputes the conclusion that such a harmonic oscillator is stable
as for every oscillatory solution, $\omega ^2$ is positive, which
means that Eq.~(\ref{eq60}) is bounded below.

In above we discussed the three examples given in Ref.~\cite{3}.
In the following we construct a new example which gives a deeper
insight into the difference between the familiar form
Eq.~(\ref{eq58}) and the correct definition of the system energy
Eq.~(\ref{eq15}), from which we can find that the stability of
system and the form of solutions to the equations of motion have a
more complicated relation in maximally nonlocal action than that
of local case. To understand it, consider the following maximally
nonlocal harmonic oscillator:
\begin{equation}
\label{eq64} L = - m\omega _0^2 q^2\cos \left( {\int\limits_{ -
T}^T {\frac{\dot {q}^2}{\omega _0^2 l_0^2 }\frac{dt}{T}} } \right)
+ \frac{1}{2}m\omega _0^2 l_0^2 G\left( {\int\limits_{ - T}^T
{\frac{\dot {q}^2}{\omega _0^2 l_0^2 }\frac{dt}{T}} } \right)\ ,
\end{equation}
\noindent where $G(t)$ is a shorthand for:
\begin{equation}
\label{eq65} G(t) = \frac{1}{2}e^t\left[ {\sin t - t\left( {\sin t
- \cos t} \right)} \right]\ .
\end{equation}
The new function $L_c $can be derived from Eq.~(\ref{eq7}):
 \bqa
\label{eq66}
 L_c &=& - m\omega _0^2 q^2\cos \left( {\int\limits_{ - T}^T {\frac{\dot
{q}^2}{\omega _0^2 l_0^2 }\frac{dt}{T}} } \right) +
\frac{1}{2}m\omega _0^2 l_0^2 G\left( {\int\limits_{ - T}^T
{\frac{\dot {q}^2}{\omega _0^2 l_0^2 }\frac{dt}{T}} }
\right)\nonumber\\
 &&+ m\dot {q}^2\int\limits_{ - T}^T
{\frac{q^2}{l_0^2 }\frac{dt}{T}\sin \left( {\int\limits_{ - T}^T
{\frac{\dot {q}^2}{\omega _0^2 l_0^2 }\frac{dt}{T}} } \right)}
 + m\dot {q}^2g\left( {\int\limits_{ - T}^T {\frac{\dot {q}^2}{\omega _0^2
l_0^2 }\frac{dt}{T}} } \right)\ ,
\eqa
 \noindent where $g(t)$ is a shorthand for:
\begin{equation}
\label{eq67} g(t) = G'(t) = e^t\left( {\cos t - t\sin t} \right)\
.
\end{equation}
By the equation of motion with respect to $L_c $ given by
Eq.~(\ref{eq8}), we can obtain:
 \be \ddot {q} + \omega ^2q = 0\ .\label{77}
  \ee
\noindent where \be \omega ^2 = \omega _0^2 \frac{\cos \left(
{\int\limits_{ - T}^T {\frac{\dot {q}^2}{\omega _0^2 l_0^2
}\frac{dt}{T}} } \right)}{\int\limits_{ - T}^T {\frac{q^2}{l_0^2
}\frac{dt}{T}\sin \left( {\int\limits_{ - T}^T {\frac{\dot
{q}^2}{\omega _0^2 l_0^2 }\frac{dt}{T}} } \right)} + g\left(
{\int\limits_{ - T}^T {\frac{\dot {q}^2}{\omega _0^2 l_0^2
}\frac{dt}{T}} } \right)}\ .\label{78}\ee
  The general
solution to Eq.~(\ref{77}) has the following form:
\begin{equation}
\label{eq68} q(t) = q_0 \cos \omega t + \frac{\dot {q}_0 }{\omega
}\sin \omega t\ .
\end{equation}
Consider a specific situation that the initial coordinate $q_0 $
is zero and the initial velocity $\dot {q}_0 $ is arbitrary:
\begin{equation}
\label{eq69} q(t) = \frac{\dot {q}_0 }{\omega }\sin \omega t\ .
\end{equation}
Substitute the solution Eq.~(\ref{eq69}) into Eq.~(\ref{78}) we
obtain:
\begin{equation}
\label{eq70} \omega ^2 = \omega _0^2 \frac{\cos \left( {\frac{\dot
{q}_0^2 }{\omega _0^2 l_0^2 }} \right)}{\frac{\dot {q}_0^2
}{\omega ^2l_0^2 }\sin \left( {\frac{\dot {q}_0^2 }{\omega _0^2
l_0^2 }} \right) + g\left( {\frac{\dot {q}_0^2 }{\omega _0^2 l_0^2
}} \right)}\ .
\end{equation}
Let $\frac{\dot {q}_0^2 }{\omega _0^2 l_0^2 }$ be $t$ and
Eq.~(\ref{eq70}) is reduced to:
\begin{equation}
\label{eq71} \omega ^2 = \omega _0^2 \frac{\cos t}{\frac{\omega
_0^2 }{\omega ^2}t\sin t + g(t)}\ .
\end{equation}
This equation can be converted to a more explicit expression:
\begin{equation}
\label{eq72} \omega ^2 = \omega _0^2 \frac{\cos t - t\sin
t}{g(t)}\ .
\end{equation}
Substitute the explicit expression of $g(t)$, Eq.~(\ref{eq67}),
into Eq.~(\ref{eq72}) we get:
\begin{equation}
\label{eq73} \omega ^2 = \omega _0^2 e^{ - t}\ ,
\end{equation}
from which we read off that for every initial value $(0,
\dot{q})$, $\omega^2$ is always positive which means an
oscillatory solution\footnote{ Such a choice of initial value has
no loss of generality, as every  general initial value $(q_0,
\dot{q}_0)$ is located on a certain closed motion curve, the
collection of which fills the space expanded by $(q_0,
\dot{q}_0)$. }. The energy of the system, defined by
Eq.~(\ref{eq15}), follows: \bqa \label{eq74}
 E &=& m\dot {q}^2\int\limits_{ - T}^T {\frac{q^2}{l_0^2 }\frac{dt}{T}\sin
\left( {\int\limits_{ - T}^T {\frac{\dot {q}^2}{\omega _0^2 l_0^2
}\frac{dt}{T}} } \right)} + m\dot {q}^2g\left( {\int\limits_{ -
T}^T {\frac{\dot {q}^2}{\omega _0^2 l_0^2 }\frac{dt}{T}} }
\right)\nonumber\\
&& + m\omega _0^2 q^2\cos \left( {\int\limits_{ - T}^T {\frac{\dot
{q}^2}{\omega _0^2 l_0^2 }\frac{dt}{T}} } \right)
 - \frac{1}{2}m\omega _0^2 l_0^2 G\left( {\int\limits_{ - T}^T {\frac{\dot
{q}^2}{\omega _0^2 l_0^2 }\frac{dt}{T}} } \right)\ .
\eqa Noting the definition of $\omega ^2$ Eq.~(\ref{78}), we
reduce Eq.~(\ref{eq74}) to:
 \bqa \label{eq75}
  E &=&
\left[ {\int\limits_{ - T}^T {\frac{q^2}{l_0^2 }\frac{dt}{T}\sin
\left( {\int\limits_{ - T}^T {\frac{\dot {q}^2}{\omega _0^2 l_0^2
}\frac{dt}{T}} } \right)} + g\left( {\int\limits_{ - T}^T
{\frac{\dot {q}^2}{\omega _0^2 l_0^2 }\frac{dt}{T}} } \right)}
\right]  \left( {m\dot {q}^2 + m\omega ^2q^2}
\right)\nonumber\\&&- \frac{1}{2}m\omega _0^2 l_0^2 G\left(
{\int\limits_{ - T}^T
{\frac{\dot {q}^2}{\omega _0^2 l_0^2 }\frac{dt}{T}} } \right)\nonumber\\
&=& 
  \frac{\omega _0^2 }{\omega ^2}\cos \left( {\int\limits_{ - T}^T
{\frac{\dot {q}^2}{\omega _0^2 l_0^2 }\frac{dt}{T}} }
\right)\left( {m\dot {q}^2 + m\omega ^2q^2} \right) -
\frac{1}{2}m\omega _0^2 l_0^2 G\left( {\int\limits_{ - T}^T
{\frac{\dot {q}^2}{\omega _0^2 l_0^2 }\frac{dt}{T}} } \right)\nonumber\\
\
.
 \eqa
  Substitute the solution Eq.~(\ref{eq69}) and
Eq.~(\ref{eq73}) into Eq.~(\ref{eq75}):
\begin{equation}
\label{eq76} E = m\omega _0^2 l_0^2 te^t\cos t -
\frac{1}{2}m\omega _0^2 l_0^2 G(t)\ .
\end{equation}
\noindent where $t$ is a shorthand for $\frac{\dot {q}_0^2
}{\omega _0^2 l_0^2 }$. Note the explicit expression of $G(t)$:
\begin{equation}
\label{eq77} E = \frac{1}{4}m\omega _0^2 l_0^2 te^t\left( {3\cos t
+ \sin t - \frac{\sin t}{t}} \right)\ .
\end{equation}
In fig. 1 we plot  qualitatively the dependence of the energy on
$t$, which illustrates that the energy of system Eq.~(\ref{eq77})
is unbounded below. Noting an important fact that the Hamiltonian
and the energy of system are the same quantity in two different
representations when the condition Eq.~(12) is satisfied, we can
infer that the maximally nonlocal harmonic oscillator
Eq.~(\ref{eq64}) is unstable, which is contrary to the conclusion
of a local case.
\begin{figure}
\begin{center}
\includegraphics[width=9cm,height=6cm]{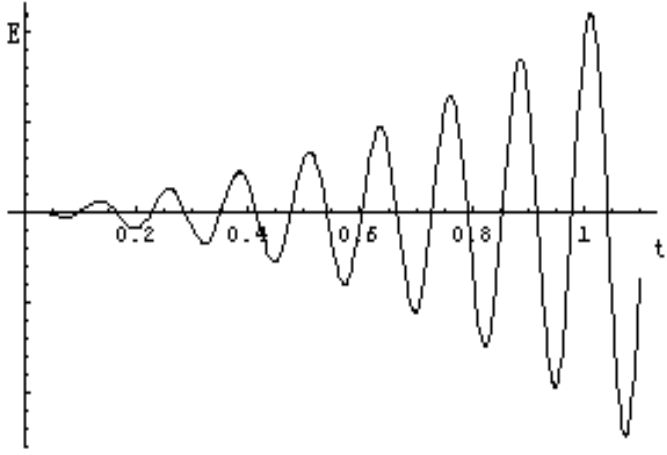}
\caption{\label{contourz}The qualitative behavior of the energy as
a function of t.}
\end{center}
\end{figure}

It is argued in Ref.~\cite{6} that a harmonic oscillator with an
unbounded below Hamiltonian is not surprising, since one can get a
simple example  by reversing the sign of the harmonic oscillator
Lagrangian:
\begin{equation}
\label{eq78} L = - \frac{1}{2}m\dot {q}^2 + \frac{1}{2}m\omega
_0^2 q^2\ ,
\end{equation}
which does not alter the oscillatory solution but the
corresponding Hamiltonian is unbounded below:
\begin{equation}
\label{eq79} H = - \frac{1}{2}m\dot {q}^2 - \frac{1}{2}m\omega
_0^2 q^2\ .
\end{equation}
However, reversing the sign of the Lagrangian, in essence, changes
the definition of energy. From Eq.~(\ref{eq15}), we can easily
obtain that the energy is defined as a conserved quantity during
the motion when the condition Eq.~(\ref{12}) is satisfied. If the
defined energy $E$ is conserved, $ - E$ is also conserved but a
stable system requires its energy $E$ (more literally, its
Hamiltonian) to be bounded below, thus meaning that arbitrary
definition of energy would lead to contradictory conclusions
concerning the stability of system.

One thing we should pay attention to is the correspondence principle which
can determine some constants that would otherwise be arbitrary in new
physics. The energy of a classical harmonic oscillator is:
\begin{equation}
\label{eq80} E = \frac{1}{2}m\dot {q}_0^2 + \frac{1}{2}m\omega
_0^2 q_0^2\ ,
\end{equation}
\noindent where $q_0 ,\dot {q}_0 $ are the initial values.
(Eq.~(\ref{eq78}) is meaningless in classical physics.) When the
energy is low, the maximally nonlocal harmonic oscillator should
be degenerated to its classical form Eq.~(\ref{eq80}). Note that
when $t = \frac{\dot {q}_0^2 }{\omega _0^2 l_0^2 }$ is a small
number, Eq.~(\ref{eq77}) is approximate to:
\begin{equation}
\label{eq81} E \approx \frac{1}{2}m\omega _0^2 l_0^2 t =
\frac{1}{2}m\dot {q}_0^2\ ,
\end{equation}
\noindent which accords with the classical form Eq.~(\ref{eq80}).
The correspondence principle justifies the definition of energy
Eq.~(\ref{eq15}) and further confirms that the maximally nonlocal
harmonic oscillatory Eq.~(\ref{eq64}), though its energy has a
classical limit, is actually unstable when the initial values are
large enough to result in deep negative energy.
\section{ Conclusions}\label{sec4}

We have established the canonical formalism of maximally nonlocal
Lagrangian by the Legendre transformation to the new function $L_c
$ (\ref{eq7}). We have also shown that when the conditions
Eq.~(\ref{eq7}) and Eq.~(\ref{12}) are satisfied, two important
conserved quantities exist, both of which can be derived from the
new function $L_c $. It is easy to conclude that the new function
$L_c $ behaves much like the Lagrangian in local theory in that
standard procedures can remain as long as the new function $L_c $
substitutes the original Lagrangian. And $L_c $ differs from the
original Lagrangian only when the functional of the dynamical
variables is introduced into the Lagrangian and Hamilton's
principle is modified to fit the maximally nonlocal Lagrangian.

The examples we have discussed above illustrate that with the new
definition of the energy of system (more literally the canonical
Hamiltonian), more complicated physical properties emerge in the
maximally nonlocal action, some of which differ drastically with
those of local cases. The difference culminates in the last
example as the positivity of the energy of a harmonic oscillatory
is demolished and the unbounded negative energy is derived from
the new definition of the energy.


\textbf{Acknowledgments}

This work was partially supported by the ``Principal Fund'' of Peking
University.

\end{document}